\documentclass[12pt]{iopart}

\usepackage[dvips]{graphicx}
\usepackage{epsfig}
\usepackage{color}
\usepackage{amssymb}
\usepackage{amsthm}

\begin{document}

\title{Field control of single x-ray photons in nuclear forward scattering}

\author{Xiangjin Kong, Wen-Te Liao and Adriana P\'alffy}

\address{Max-Planck-Institut f\"ur Kernphysik, Saupfercheckweg 1, D-69117 Heidelberg, Germany}

\ead{palffy@mpi-hd.mpg.de}

\begin{abstract}
Means to coherently control single x-ray photons in resonant scattering of light off nuclei  by electric or magnetic fields  are investigated theoretically. 
In order to derive the time response in nuclear forward scattering, we adapt the Maxwell-Bloch equations known from quantum optics to describe the resonant light pulse propagation through a nuclear medium. 
Two types of  time-dependent perturbations of nuclear forward scattering are considered for coherent control of the resonantly scattered x-ray quanta. First, the simultaneous coherent propagation  of two pulses through the nuclear sample is addressed. We find that the signal of a weak pulse can be enhanced or suppressed by a stronger pulse simultaneously propagating through the sample in counter-propagating geometry. 
Second, the effect of a time-dependent hyperfine splitting is investigated and we put forward a scheme that allows parts of the spectrum to be shifted forward in time. 
This is the inverse effect of coherent photon storage and may become a valuable technique if single x-ray photon wavepackets are to become the information carriers in future photonic circuits.

\end{abstract}





\maketitle
Recent experimental developments of coherent light sources have opened the x-ray parameter regime for fascinating coherent control concepts originally developed in  quantum optics. Thus, new fields such as  x-ray quantum optics \cite{adams2013} and nuclear quantum optics \cite{Buervenich2006} emerge. The interest in nuclear systems is sustained by  the recent commissioning of  X-ray Free Electron Laser (XFEL) facilities \cite{slac,Sacla,Emma2010.NP,Gutt2012.PRL,Ishikawa2012.NP} and the development of x-ray optics devices \cite{SchroerAPL,SchroerPRL,KangAPL, Shvydko2010,Shvydko2011,mimura2013,Osaka2013}  which bring into play higher photon  frequencies. Nuclei with low-lying collective states therefore become candidates for nuclear quantum optics \cite{Buervenich2006,Coussement2002,wong2011,Olga2013a,Olga2013b} or nuclear coherent population transfer \cite{Liao2011,Liao2013}. 

Coherent control tools based on nuclear cooperative effects \cite{Van1999, shvydko2000, Roehlsberger2004, Roehlsberger2010,Roehlsberger2012} are known also from nuclear forward scattering (NFS) experiments with third-generation  light sources. The underlying physics here relies on the delocalized nature of the nuclear excitation produced by coherent XFEL or synchrotron radiation (SR) light, i.e., the formation of  so-called nuclear excitons.  For instance, a NFS setup in planar thin film waveguides \cite{Ralf2004}  was used for novel quantum optics experiments in the x-ray regime using nuclei instead of atoms. The excitonic nature of the nuclear excitation in NFS was 
exploited to identify the cooperative Lamb shift \cite{Roehlsberger2010}, or to demonstrate electromagnetically induced transparency \cite{Roehlsberger2012} and spontaneously generated coherence \cite{Evers2013} in a nuclear system. 
 Furthermore, NFS setups also offer a framework for control of single x-ray photons, which might become a useful tool for optics and quantum information applications at shorter wavelengths on the way 
towards more compact  photonic devices \cite{Politi2008}. Phase-sensitive storage and $\pi$ phase modulation for single hard x-ray photons in a NFS setup have been recently proposed \cite{Liao2012a}, as well as the generation of a nuclear polariton with two entangled counter-propagating branches \cite{Liao2013b} comprising a single x-ray photon. Using M\"ossbauer sources, the coherent control of the single-photon wavepackets shape has been recently demonstrated \cite{Olga2013b}.

In this work we focus on advanced field-control means to coherently manipulate the resonant x-ray pulse propagation through  a nuclear medium and the corresponding single-photon wave packets. In particular, we first consider the case of two resonant pulses simultaneously propagating through the same nuclear sample. A counter-propagating geometry is envisaged in order to easily discern between the scattered signal of the individual pulses. We find that the signal of a weak pulse (potentially a single-photon wavepacket) can be enhanced or suppressed by the presence of a counter-propagating stronger pulse, depending on the corresponding time delay. The underlying mechanisms of this behaviour are identified and discussed. The signal enhancement and suppression effects in the interaction between the two pulses might prove very useful for enhancing detection and control of the single-photon wave packet.
 Secondly, we address the
 effect of time-dependent hyperfine magnetic fields that switch the nuclear system from  the degenerate, two-level system case, to a non-degenerate multi-hyperfine-level one. External magnetic fields have been used to control the  NFS  and in particular to store the nuclear excitation \cite{Shvydko1996,Palffy2009,Liao2012a,Liao2012b}. As a new feature, we discuss  here a magnetic field control sequence which allows the shift of the NFS signal  forward in time, i.e., towards shorter, earlier times. 
This is the inverse effect of coherent photon storage presented in Ref. \cite{Liao2012a} which shifts the NFS signal towards later times. Using these two magnetic-field switching techniques, one has efficient time-signal processing tools of single-photon wave packets.

In order to study the field-control effects described above, a versatile theoretical method is required which allows to easily incorporate perturbations of the NFS signal by means of time-dependent  electric or magnetic fields. 
There are a number of theoretical approaches to treat the coherent nuclear excitations induced by SR pulses and calculate the amplitude of the scattered light \cite{Hannon1999}. The first  time-dependent theory of NFS of SR was developed by Kagan, Afanas’ev, and Kohn \cite{Kagan1979}.  Fourier transformation from the frequency to the time domain as shown in Ref.~\cite{Kagan1979} has been  used ever since in many works  to consider more complicated cases of interactions of nuclei with their environment \cite{Hannon1999,Sturhahn1994,Smkohn1995}. Alternatively,  the scattering problem can be directly treated semi-classically in time and space, see Ref.~\cite{Shvydko1999N}, leading 
to a wave equation for the scattered field to be solved iteratively. In this work we adopt a more general approach from atomic quantum optics based on the  Maxwell-Schr\"odinger or Maxwell-Bloch equations (MBE) \cite{Scully2006}. This allows to determine the  field propagation through the nuclear medium easily taking into account additional perturbation such as time-dependent magnetic fields or the simultaneous propagation of several light pulses through the same sample. 
 The parameter regime for which the MBE reproduce the well-known dynamical beat results for a single nuclear transition is deduced. The case of NFS off multi-level nuclei is discussed and the form of the MBE is derived  taking into account hyperfine splitting for the case of the  $^{57}\mathrm{Fe}$ M\"ossbauer nucleus.  Using a forward-backward decomposition, the MBE can also be generalized to treat the propagation and medium response of two counter-propagating pulses. For atomic resonant media, two-pulse propagation in short-pulse electromagnetically-induced transparency (EIT) scenarios have been previously successfully described using the MBE formalism \cite{Eberly2007}.

The paper is organized as follows. In Sec. \ref{mbe} we derive the MBE for the scattering of light off identical nuclei and discuss the parameter regime for which they describe the NFS spectra. 
The case of field-controlled NFS with two pulses simultaneously propagating in the same nuclear sample is presented in Sec. \ref{elcontrol}.  Sec. \ref{magncontrol} addresses forwarding the nuclear response in time by means of 
 time-dependent external hyperfine magnetic fields. Finally, Sec.~\ref{Conclusions}  summarizes the results.

\section{Theoretical Approach \label{mbe}}
\begin{figure}[htbp]
\centering
\includegraphics[width=\textwidth]{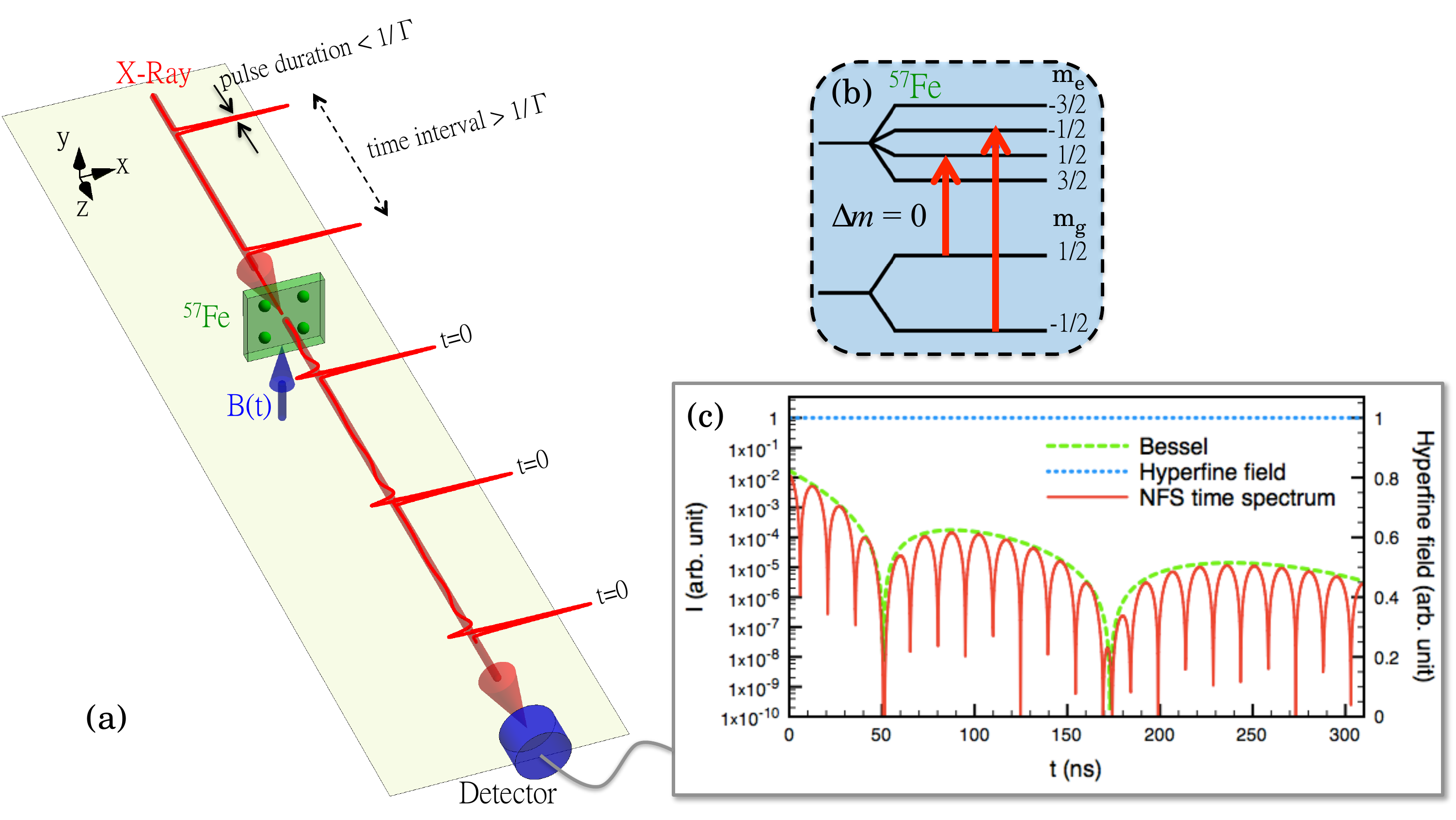}
\caption{\label{fig1} (a) Typical NFS setup. The resonant x-ray pulse shines perpendicularly to the nuclear sample depicted by the green rectangle. After each pulse, the delayed nuclear response in the forward direction is recorded by the detector. The blue thick vertical arrow shows the applied magnetic field  {\bf B}(t). (b)  $^{57}$Fe ground and first excited state nuclear hyperfine levels. In this example, the $\Delta m=0$ transitions are driven by linearly polarized x-rays.  (c)  Intensity of the coherently scattered light in the forward direction (red solid line) for an incident field driving the $\Delta m=0$ transitions. The envelope given by the Bessel function for the degenerate states case is shown by the green long-dashed line. The hyperfine magnetic field depicted by the blue short-dashed line is kept constant during the scattering for this example. }
\end{figure}

In a typical NFS experiment, monochromatized light pulses shine perpendicular to a sample containing M\"ossbauer nuclei, usually  $^{57}\mathrm{Fe}$. The delayed nuclear response is then recorded by observing the resonantly scattered light in the forward direction, as illustrated schematically in Fig.~\ref{fig1}(a). 
The interval between successive light pulses is chosen long enough to facilitate the nuclear response detection, typically larger than $1/\Gamma$, where $\Gamma$ denotes the nuclear spontaneous decay rate.
The driven magnetic dipole (M1) nuclear transition connects the  $^{57}\mathrm{Fe}$ ground state characterized by spin $I_g=1/2$ to the first excited state at 14.413 keV with  $I_e=3/2$. 
The hyperfine-split level scheme of $^{57}\mathrm{Fe}$ for the states of interest is depicted in Fig.~\ref{fig1}(b). The resonant scattering occurs via an  excitonic state, i.e., an excitation coherently spread out over a large number of nuclei.  In case of coherent scattering, the nuclei return to their initial state, such that the scattering path and the number of occurred events  are unknown. This leads to cooperative emission, with scattering only in forward direction (except for the case of Bragg scattering~\cite{Kagan1979,Hannon1999,Smirnov1996,Smirnov1996NE}) and decay rates modified by the formation of sub- and superradiant states as key signatures. The observed decay signal is therefore far from being exponential, as can be seen in the example presented in Fig.~\ref{fig1}(c).

The exciton picture \cite{Hannon1999,Roehlsberger2004,Scully2006b} justifies the coherently scattered radiation proceeding in the forward direction, but does not provide a straightforward manner to correctly derive the scattering spectrum. This can be rather achieved by means of the wave equation for the time-dependent field propagation which reveals the field intensity at the exit  from the sample. The ansatz of forward emission of the resonantly scattered light 
is however related to the exciton picture and enters the MBE phenomenologically. 

In quantum optics, the light-nuclei interaction is typically described by monitoring the quantum time evolution of the density operator $\hat{\rho}$, given by the master equation \cite{Scully2006}
\begin{equation}
\partial_t\hat{\rho}=\frac{1}{i\hbar}[\hat{H},\hat{\rho}] + \hat{\rho}_s\, .
\end{equation}
Here, $\hat{H}$ is the interaction Hamiltonian between the
matter and the incident electromagnetic field and $\hat{\rho}_s$ describes 
decoherence processes such as spontaneous decay.
For a two-level system corresponding to a single nuclear resonance with ground state $|g\rangle$ and excited state $|e\rangle$, the interaction Hamiltonian is given by 
\begin{equation}
\hat{ H}=-\frac{\hbar}{2}\left(\begin{array}{cc}
  0  & \Omega_{p}^* \\
  \Omega_{p}  & 2 \Delta_{p}
\end{array} \right)\, ,
\end{equation}
where $\hbar$ is the reduced Planck constant, and $\Delta_p$ is the detuning (i.e., mismatch) between the field and nuclear transition frequencies. Furthermore,
$\Omega_p$ denotes the Rabi frequency defined as
\begin{equation}
 \Omega_p=\frac{1}{\hbar}\langle e |\hat{H}| g\rangle \, .
\end{equation}
By using  the Coulomb gauge for the vector potential $\vec{A}(z,t)$ and the rotating wave approximation, we can obtain a useful expression of the  reduced interaction matrix element,
\begin{eqnarray}
 \langle e | \hat{H}| g\rangle &=& -\langle e |\hat{\vec{j}}(\vec{k}) \cdotp\hat{\vec{A}}(z,t)| g\rangle \\
                                   &=& E(z,t)\sqrt{2\pi}\sqrt{\frac{L+1}{L}}\frac{k^{L-1}}{(2L+1)!!} \sqrt{B(\varepsilon/\mu L,|g\rangle\rightarrow |e\rangle)} \nonumber \\
					&\equiv &  E(z,t) \alpha (\varepsilon/\mu L,|g\rangle\rightarrow |e\rangle) \, , \label{rabi}
\end{eqnarray}
where $\hat{\vec{j}}(\vec{k})$  is the current density operator in momentum representation, 
$E(z,t)$ is the electric field envelope,
$L$ is the angular momentum of the transition, $\varepsilon/\mu$ the transition type (electric/magnetic), and  $B(\varepsilon/\mu L,|g\rangle\rightarrow |e\rangle)$  the nuclear reduced transition probability \cite{ring1980}. 
For the equation above we have considered the case of a single nuclear transition from a degenerate ground state.
Typically, in atomic quantum optics only electric dipole transitions are of interest and $\alpha (\varepsilon 1,|g\rangle\rightarrow |e\rangle)$ stands then for the electric dipole moment. In our case, we have written in Eq.~(\ref{rabi}) the general expression of the Rabi frequency involving the electromagnetic multipole moment $\alpha (\varepsilon/\mu L,|g\rangle\rightarrow |e\rangle)$.

With the notation $\rho_{mn}=\langle m | \hat{\rho} | n\rangle$ with $\{m,n\}\in\{e,g\}$ we obtain the Bloch equations
\begin{eqnarray}
\partial _t\rho _{gg}&=&\Gamma\rho _{ee}-\frac{i}{2}(\Omega_{p}\rho _{ge}-\Omega_{p}^*\rho_{eg})\, ,\nonumber \\
\partial _t\rho _{eg}&=&-\left(i\Delta_p +\frac{\Gamma}{2}\right)\rho _{eg}-\frac{i}{2}\Omega_{p}(\rho_{ee}-\rho _{gg})\, ,\nonumber \\
\partial _t\rho _{ee}&=&-\Gamma\rho _{ee}+\frac{i}{2}(\Omega_{p}\rho _{ge}-\Omega_{p}^*\rho_{eg})\, ,
\label{bloch}
\end{eqnarray}
where the spontaneous decay rate $\Gamma$  comprises the radiative and the internal conversion channel.

By coupling the equations above for
the density matrix to the Maxwell wave equation,  we can describe the dynamics of both matter and radiation field, i.e., the propagation of a light pulse through the  resonant medium taking into account also the sample response. In the following we consider an electromagnetic wave with the polarization vector $\vec{e}_{x}$, frequency $\omega$ and wave number $k_0=\omega/c$ (here $c$ denotes the speed of light) with a slowly varying envelope 
\begin{equation}
\vec{E}(z,t)=E(z,t)e^{-i(\omega t-k_{0}z)}\vec{e}_{x} \, .
\end{equation}
Considering only unidirectional propagation in the forward direction according to our ansatz,  the wave equation 
\begin{equation}
\left(\frac{\partial^2}{\partial z^2}-\frac{1}{c^{2}}\frac{{\partial}^{2}}{\partial{t}{^2}}\right)\vec{E}(z,t)=\frac{4\pi}{c^2}\frac{\partial}{\partial t}\vec{I}(z,t)
\end{equation}
for the electric field intensity   has as source term the macroscopic current density
$\vec{I}(z,t)$ induced by the radiation in the system of resonant nuclei. The induced current density can be written as 
\begin{equation}
 \vec{I}(z,t)=I(z,t)e^{-i(\omega t-k_0z)}\vec{e}_{x}\, .
\end{equation}
We consider the parameter regime for which $|\frac{\partial E(z,t)}{\partial t}|,\ | c\frac{\partial E(z,t)}{\partial z}|\ll|\omega E(z,t)|$ holds.  In the slowly varying envelope approximation, the wave equation reduces to
\begin{equation}
\frac{\partial E(z,t)}{\partial z}+\frac{1}{c}\frac{\partial E(z,t)}{\partial t}=-\frac{2\pi}{c} I(z,t)\, .
\label{max}
\end{equation}
The crucial step here is to express the current density with the help of the density matrix in order to couple the Bloch and Maxwell equations. For a two-level system interacting with the field in atomic quantum optics, the current can be expressed with the help of the coherence $\rho_{eg}$ and the dipole moment $\alpha (\varepsilon 1,|g\rangle\rightarrow |e\rangle)$. Following the argument presented in Ref.~\cite{Shvydko1999N}, 
the current density  for a single nuclear resonance is obtained by summing over all nuclei participating in the coherent scattering and tracing over $\hat{\vec{j}}(\vec{k})e^{ik_0z}\hat{\rho}$. Taking into account the alternative form of the Hamiltonian with the vector potential written in the Coulomb gauge, $\hat{H}=i\hat{\vec{j}}(\vec{k})\cdot \vec{e}_xe^{ik_{0}z}E(z,t)/\omega$, we can relate to the matrix element in Eq. (\ref{rabi}) and express the current in the simplified form
\begin{eqnarray}
 I(z,t)&=&N\langle e |\hat{\vec{j}}(\vec{k})e^{ik_{0}z} | g\rangle \rho_{eg} \\
       &=&\frac{\omega}{i}N\alpha (\varepsilon/\mu L,|g\rangle\rightarrow |e\rangle)\rho_{eg}\, ,
\label{i}
\end{eqnarray}
where $N$ is the particle number density and we take into account all nuclei over which the excitation is coherently shared. Combining Eqs. (\ref{rabi}), (\ref{max}) and (\ref{i}) we obtain an additional equation involving the Rabi frequency,  
\begin{equation}
\frac{1}{c}\partial _t\Omega_{p}(z,t)+\partial _z\Omega_{p}(z,t)=i\frac{2\pi\omega N[\alpha (\varepsilon/\mu L,|g\rangle\rightarrow |e\rangle)]^2}{\hbar c}\rho_{eg}\, .
\end{equation}
Together with the three Bloch equations (\ref{bloch}), we now have arrived at the MBE for the Rabi frequency. The scattered field is then proportional to $ \Omega_p$ and the scattered intensity $I\propto |\Omega_p|^2$. We proceed now with some changes of notation in order to facilitate the comparison with established NFS results. The expression of the radiative nuclear  decay rate $\Gamma_{\gamma}$ is also connected to the reduced transition probabilities 
$B(\varepsilon/\mu L,|e\rangle\rightarrow |g\rangle)$ via 
\begin{equation}
\Gamma_{\gamma}=\frac{8\pi(L+1)}{L[(2L+1)!!]^2}\left(\frac{E_0}{\hbar c}\right)^{2L+1}B(\varepsilon/\mu L,|e\rangle\rightarrow |g\rangle)\, ,
\label{gamma}
\end{equation}
where $E_0$ denotes  the transition energy and
\begin{equation}
B(\varepsilon/\mu L,|e\rangle\rightarrow |g\rangle)=\frac{2I_g+1}{2I_e+1}B(\varepsilon/\mu L,|g\rangle\rightarrow |e\rangle)\, ,
\end{equation}
i.e., they are equal when considering the case of a single nuclear resonance. The resonant cross section can also be expressed as 
\begin{equation}
 \sigma=\frac{2\pi}{k_0^2}\frac{2I_e+1}{2I_g+1}\frac{\Gamma_{\gamma}}{\Gamma} 
=[\alpha (\varepsilon/\mu L,|g\rangle\rightarrow |e\rangle)]^2\frac{8\pi k}{\hbar\Gamma} \, .
\end{equation}
Introducing the dimensionless effective thickness \cite{Shvydko1999N} $\xi=N\sigma L/4$  with $L$ the length of the sample, we can rewrite the wave equation in the MB equations as
\begin{equation}
\frac{1}{c}\partial _t\Omega_{p}(z,t)+\partial _z\Omega_{p}(z,t)=i\eta\rho _{eg}(z,t)\, ,
\label{maxnice}
\end{equation}
with  $\eta=\frac{\xi\Gamma}{L}$.

As initial conditions for the MBE we now consider
\begin{eqnarray}
\rho_{mn}(z,0)&=&\delta_{mg}\delta_{ng}\, , \nonumber \\
\Omega_{p}(z,0)&=&0\, , \nonumber \\
\Omega_{p}(0,t)&=&\Omega_{0p}\delta(t-\tau) \, ,
\label{incond}
\end{eqnarray}
where $\tau$ marks the arrival of the incident resonant light pulse. In the following we set the detuning $\Delta_p$ to zero. Taking the incident pulse as a small perturbation such that $\Omega_{p}\ll\Gamma$ and no Rabi oscillations may occur, we obtain in first order perturbation theory from Eqs. (\ref{bloch}) and (\ref{maxnice}) only two coupled equations for $\Omega_{p}$,
\begin{eqnarray}
\partial _t\rho _{eg}=-\frac{\Gamma}{2}\rho _{eg}+\frac{i}{2}\Omega_{p}\, ,\nonumber \\
\frac{1}{c}\partial _t\Omega_{p}+\partial _z\Omega_{p}=i\eta\rho _{eg}\, .
\end{eqnarray}
Performing a change of variable and using the Fourier transform, the dispersion relation of the system
can be obtained \cite{LiaoBook}, 
\begin{equation}
k(\omega)=\frac{\omega}{c}-\frac{\eta}{2\omega}-i\frac{\Gamma}{2c}\, .
\end{equation}
The solution for the Rabi frequency can be found by inverse Fourier transform
\begin{eqnarray}
\Omega_p(z,t)&=& \frac{1}{2\pi}e^{-\frac{\Gamma}{2}[\frac{z}{c}+(t-\tau)]}\int_{-\infty}^{\infty}e^{-i[(\frac{\omega}{c}-\frac{\eta}{2\omega})z-\omega(t-\tau)]}d\omega  \\
&=&\left\{\delta\left[\frac{z}{c}-(t-\tau)\right] -
\frac{\xi\Gamma z}{L}\frac{J_1\left[2\sqrt{(\frac{\xi\Gamma z}{L})(t-\tau-\frac{z}{c})}\right]}{2\sqrt{(\frac{\xi\Gamma z}{L})(t-\tau-\frac{z}{c})}}\right\}e^{-\frac{\Gamma}{2}(\frac{z}{c}+t-\tau)}\, ,\nonumber
\label{dbeat}
\end{eqnarray}
where  $J_1(z)$ is the Bessel function of the first kind. The terms $z/c$ are typically negligible because  $L/c$ is much smaller than $(t-\tau)$. With this, the result above reproduces the expression of the dynamical beat \cite{Kagan1979,Kagan1999,Hannon1999,Shvydko1999N} known from the time-dependent theory of NFS for a single nuclear resonance. An illustration of the dynamical beat for a test case is given by the green dashed line in Fig. \ref{fig1}(c). We would like to emphasize here that the dynamical beat is a general feature for the propagation of short weak laser pulses through resonant matter and by no means limited to NFS, as shown by earlier studies in atomic systems \cite{Burnham1969,Crisp1970,Lamb1971}.

The MBE become more complicated for the case of the resonant driving of several nuclear resonances in a hyperfine-split, multi-level system. The typical example is  ${^{57}}\mathrm{Fe}$ in a hyperfine magnetic field which has two ground   ($I_g=1/2$)  and four excited  ($I_e=3/2$)  magnetic sublevels. The hyperfine levels are coupled by six transitions, depending on the magnetic field geometry and polarization of the incident SR or XFEL field. 
Let us first consider the x-ray pulse is linearly polarized and the direction of polarization is parallel to the $x$ axis. The magnetic field {\bf B}(t) that sets the quantization axis for the nuclear ground and excited state spin projections  $m_g$ and $m_e$ is parallel to the $y$ axis, as depicted in Fig.~\ref{fig1}(a). 
In this scenario, the two $\Delta m=m_e-m_g=0$ magnetic dipole transitions will be driven by the incident pulse. The MBE include then a number of Clebsch-Gordan coefficients that quantify the individual couplings between the four states,
\begin{eqnarray}
\partial _t\rho _{11}&=&\Gamma(C_{14}^2\rho _{44}+C_{15}^2\rho _{55})
-\frac{i}{2}C_{15}(\Omega_{p}\rho _{15}-\Omega_{p}^*\rho_{51})\,  , \nonumber \\
\partial _t\rho _{22}&=&\Gamma(C_{24}^2\rho _{44}+C_{25}^2\rho _{55})
-\frac{i}{2}C_{24}(\Omega_{p}\rho _{24}-\Omega_{p}^*\rho_{42})\,  , \nonumber \\
\partial _t\rho _{42}&=&-\frac{1}{2}(2i\Delta_{p,4\rightarrow2}+C_{14}^2\Gamma+C_{24}^2\Gamma)\rho _{42}
-\frac{i}{2}C_{24}\Omega_{p}(\rho_{44}-\rho _{22})\,  , \nonumber \\
\partial _t\rho _{44}&=&-\Gamma(C_{14}^2+C_{24}^2)\rho _{44}
+\frac{i}{2}C_{24}(\Omega_{p}\rho _{24}-\Omega_{p}^*\rho_{42})\,  , \nonumber \\
\partial _t\rho _{51}&=&-\frac{1}{2}(2i\Delta_{p,5\rightarrow1}+C_{15}^2\Gamma+C_{25}^2\Gamma)\rho _{51}
-\frac{i}{2}C_{15}\Omega_{p}(\rho_{55}-\rho _{11})\,  , \nonumber \\
\partial _t\rho _{55}&=&-\Gamma(C_{15}^2+C_{25}^2)\rho _{55}
+\frac{i}{2}C_{15}(\Omega_{p}\rho _{15}-\Omega_{p}^*\rho_{51})\,  , \nonumber \\
\frac{1}{c}\partial _t\Omega_{p}&+&\partial _z\Omega_p=i\eta'(a_{51}\rho _{51}+a_{42}\rho _{42})\, .
\label{hfseqs}
\end{eqnarray}
In the above equations, the states $|1\rangle$ and $|2\rangle$ denote the  two ground states with $m_g=1/2$ and $m_g=-1/2$, respectively, and $|3\rangle$, $|4\rangle$, $|5\rangle$ and $|6\rangle$  the four excited states with  $m_e=-3/2$,  $m_e=-1/2$, $m_e=1/2$ and $m_e=3/2$, respectively. The shortened notation used for the Clebsch-Gordan coefficients \cite{Edmonds} is $C_{ij}=C(I_g\,I_e\,  1;m_g\,  m_e\, M )$ where $i\in \{1,2\}$ sets the value of $m_g$ and $j\in \{3,4,5,6\}$ the one of $m_e$. Furthermore, $\Delta_{p,4\rightarrow2}=\omega_{42}-\omega$ and $\Delta_{p,5\rightarrow1}=\omega_{51}-\omega$, where $\omega_{51}$ and $\omega_{42}$ are the resonant frequencies of the $|1\rangle\rightarrow|5\rangle$ and $|2\rangle\rightarrow|4\rangle$ transitions, respectively. The coefficients $\eta'$, $a_{51}$ and $a_{42}$ can be deduced by studying the limiting case when the magnetic field {\bf B}(t) goes to zero and Eqs. (\ref{hfseqs}) should resume the form of (\ref{bloch}) and (\ref{maxnice}).  The last equation 
in (\ref{hfseqs}) then  becomes
\begin{equation}
\frac{1}{c}\partial _t\Omega_{p}+\partial _z\Omega_p=i\eta\left(\frac{\rho _{51}}{C_{15}}+\frac{\rho _{42}}{C_{24}}\right)\, .
\end{equation} 
The MBE is therefore a very convenient method to treat NFS involving multiple resonances since the system of equations can be solved numerically. 
For completion, the corresponding equations for the case of a circularly  polarized pulse driving the four $\Delta m=m_e-m_g=\pm1$ transitions between the six ground and excited  hyperfine levels are given in the Appendix.

Comparison of theoretical and experimental NFS  results for SR show very good agreement. This might appear as surprising since most theoretical approaches, including the MBE discussed here, rely on the classical Maxwell equation for the scattered field. However, in experiments the produced excitation is very weak, such that typically either no photon or one photon is resonantly scattered per pulse and the spectra describe the propagation of a single-photon wavepacket. The legitimate question may arise how come does the classical field correctly describe the behavior of single photons? This would be the case if the photon state under investigation  were a coherent state \cite{Scully2006}. In our case, the weak excitation produced by SR pulses can be described by the coherent-like state  $C_0|0\rangle + C_1|1\rangle +C_2|2\rangle + \ldots$ where $|n\rangle$ is the $n$-photon  Fock state and $|C_0|^2\gg|C_1|^2\gg|C_2|^2\gg\ldots$. This relation between the observed photon number events for small $n$ is verified by typical NFS experiments and justifies our classical field treatment for single photons. A rigorous quantum treatment of NFS will hopefully provide more insight in the behaviour of single and few x-ray photons in nuclear samples.

\section{Two resonantly  propagating pulses \label{elcontrol}}
\begin{figure}[htbp]
\centering
\includegraphics[width=0.40\textwidth]{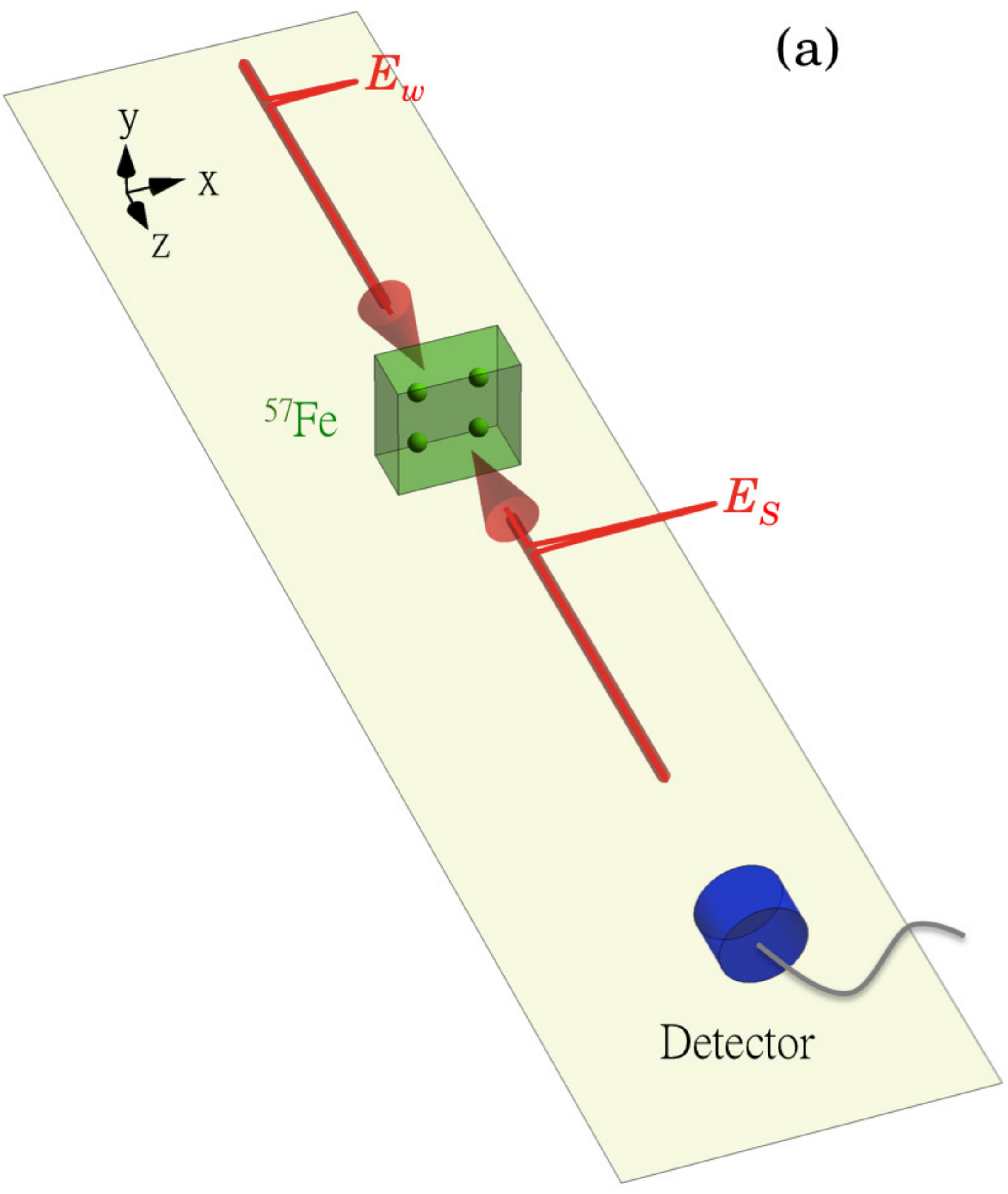}
\includegraphics[width=0.55\textwidth]{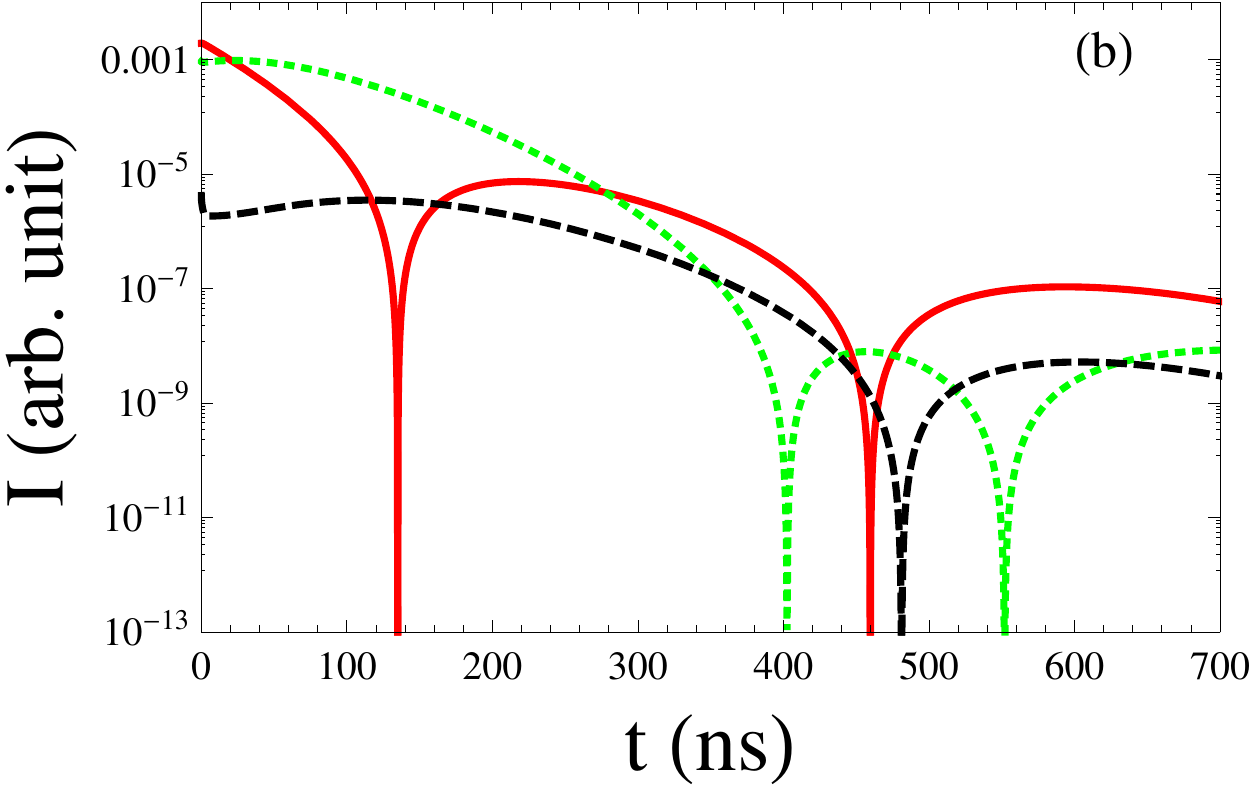}
\caption{\label{fig2} (a) Counter-propagating pulses setup with the strong pulse reaching the sample prior to the weak pulse. 
(b) NFS time spectra $I\propto|\Omega _{w}|^2$ for the weak pulse in the absence (red solid line) or presence of a stronger counter-propagating pulse $\Omega _{s}$. The latter  reaches the sample prior to the weak pulse. The time delay $\Delta\tau=\tau_s-\tau_w$ between the two pulses is $-10$ ns (green dotted line) and $-50$ ns (black  dashed line). The time origin is set by the center of the incident weak pulse reaching the sample at $z=0$.}
\end{figure}

Let us consider the case of two resonant pulses interacting simultaneously with a nuclear target containing ${^{57}}\mathrm{Fe}$ M\"ossbauer nuclei.
We choose the counter-propagating geometry as shown schematically in Fig. \ref{fig2}(a) such that the two signals can be easily separated experimentally. The recent development of normal-incidence x-ray mirrors \cite{Shvydko2010,Shvydko2011} is an important step allowing such more complex setup geometries. 
For simplicity we assume  a single nuclear transition resonant with the two light pulses which reach the target from opposite directions at $z=0$ and $z=L$. We consider the case of two pulses both with zero detuning $\Delta$ but of different intensity. A weak pulse of Rabi frequency $\Omega _{w}$  is perturbed and controlled by the simultaneous passage of a stronger pulse $\Omega _{s}$ through the sample. 
The physical case behind  such a setup may involve a weaker pulse which produces a single-photon excitation that can in turn be controlled by a more intense XFEL pulse.   In order to describe the fields in the counter-propagating geometry we consider a backward-forward decomposition of the  radiation field \cite{lin2009},
\begin{equation}
\vec{E}(z,t)=E_{w}(z,t)e^{-i(\omega t-k_{0}z)}\vec{e_{x}} +E_{s}(z,t)e^{-i[\omega t-k_{0}(L-z)]}\vec{e_{x}}\, .
\end{equation}
In our case, since for each pulse only the respective forward scattering wave is taken into account,  each term in the equation above represents the contribution of  one of the  pulses. For the numerical calculation we use the same decomposition also for the coherence terms
\begin{equation}
\rho_{eg}(z,t)=\rho_{egw}(z,t)e^{-i(\omega t-k_{0}z)}
+\rho_{egs}(z,t)e^{-i[\omega t-k_{0}(L-z)]}\, ,
\end{equation}
and the Rabi frequencies, 
\begin{equation}
\Omega(z,t)=\Omega_{w}(z,t)+\Omega_{s}(z,t)\, .
\end{equation}
A similar decomposition in the MBE was used to describe the coherent propagation of Stokes light  in a $\Lambda$ three-level amplifier, where the Raman and fluorescence components play the role of the two counter-propagating in our setup \cite{Eberly1989}. 
Writing separately the wave equations for the forward and backward Rabi frequencies, we obtain the MBE
\begin{eqnarray}
\partial _t\rho _{ee}&=&-\Gamma\rho _{ee}+\frac{i}{2}\left[(\Omega _{w}\rho _{gew}-\mathrm{c.c.})+ \right.
(\Omega _{s}\rho _{ges}
- \mathrm{c.c.})
\nonumber \\
&+&(\Omega _{w}\rho _{ges}e^{-ik_{0}L+i2k_{0}z}-\mathrm{c.c.})
+\left.(\Omega _{s}\rho _{gew}e^{ik_{0}L-i2k_{0}z}-\mathrm{c.c.})\right]\, ,
\nonumber \\
\partial _t\rho _{gg}&=&\Gamma\rho _{ee}-\frac{i}{2}[(\Omega _{w}\rho _{gew}-\mathrm{c.c.})+
(\Omega _{s}\rho _{ges}-\mathrm{c.c.})
\nonumber \\
&+&(\Omega _{w}\rho _{ges}e^{-ik_{0}L+i2k_{0}z}-\mathrm{c.c.})
+(\Omega _{s}\rho _{gew}e^{ik_{0}L-i2k_{0}z}-\mathrm{c.c.})]\, ,
\nonumber \\
\partial_t\rho_{egw}&=&-\left(i\Delta+\frac{\Gamma}{2}\right)\rho _{egw}-\frac{i}{2}\Omega _{w}(\rho _{ee}-\rho _{gg})\, ,
\nonumber \\
\partial_t\rho_{egs}&=&-\left(i\Delta+\frac{\Gamma}{2}\right)\rho _{egs}-\frac{i}{2}\Omega _{s}(\rho _{ee}-\rho _{gg})\, ,
\nonumber \\
&&\frac{1}{c}\partial _t\Omega _{w}+\partial _z\Omega _{w}=i\eta\rho _{{egw}}\, ,
\nonumber \\
&&\frac{1}{c}\partial _t\Omega _{s}-\partial _z\Omega _{s}=i\eta\rho _{{egs}}\, ,
\label{mbcounter}
\end{eqnarray}

The MBE above can be solved numerically. For numerical efficiency, we consider instead of  incident delta pulses in Eq. (\ref{incond}) a Gaussian pulse shape $\Omega(z,t)=\Omega_0 e^{-\frac{(t-\tau)^2}{\sigma^2}}$  with $\sigma=1$ ns, which is still much shorter than the nuclear decay time scale of hundreds of ns (the nuclear spontaneous decay rate, including both the radiative and the internal conversion channels, is $\Gamma=1/141$ GHz). As numerical example, the weak pulse with initial Rabi frequency  $\Omega _{w0}=\Gamma/10$ reaches the sample ($z=0$) at $\tau_w$  in the presence of a stronger pulse  ($\Omega _{s0}=200\Gamma$) arriving at other end of  the sample ($L=10$~$\mu$m) at $\tau_s$ with positive or negative time delay and propagating through the sample in the opposite direction.
The effective thickness of the sample was chosen $\xi=15$.   The results for positive and negative time delay are presented in Figs. \ref{fig2} and \ref{fig3}.

We see that the presence of the stronger pulse plays an important role on the propagation of the weaker resonant pulse. We address the two situations of positive and negative pulse delay separately.

\subsection{$\Delta \tau<0$}
\begin{figure}[htbp]
\centering
\includegraphics[width=0.6\textwidth]{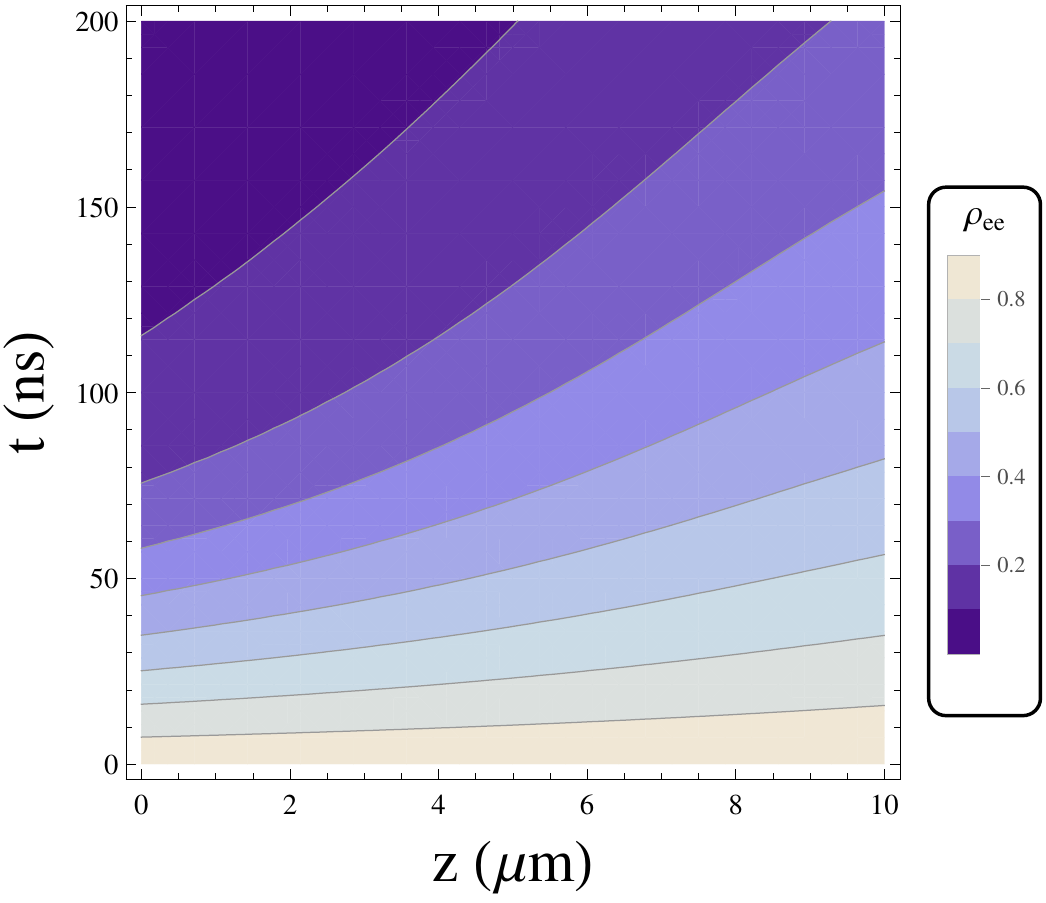}
\caption{\label{fig4} Excited state population $\rho_{ee}$ produced solely by the strong pulse as a function of time (here $t=0$ denotes the center of the strong pulse entering the sample) and position $z$ in the sample. }
\end{figure}

\begin{figure}[htbp]
\centering
\includegraphics[width=0.6\textwidth]{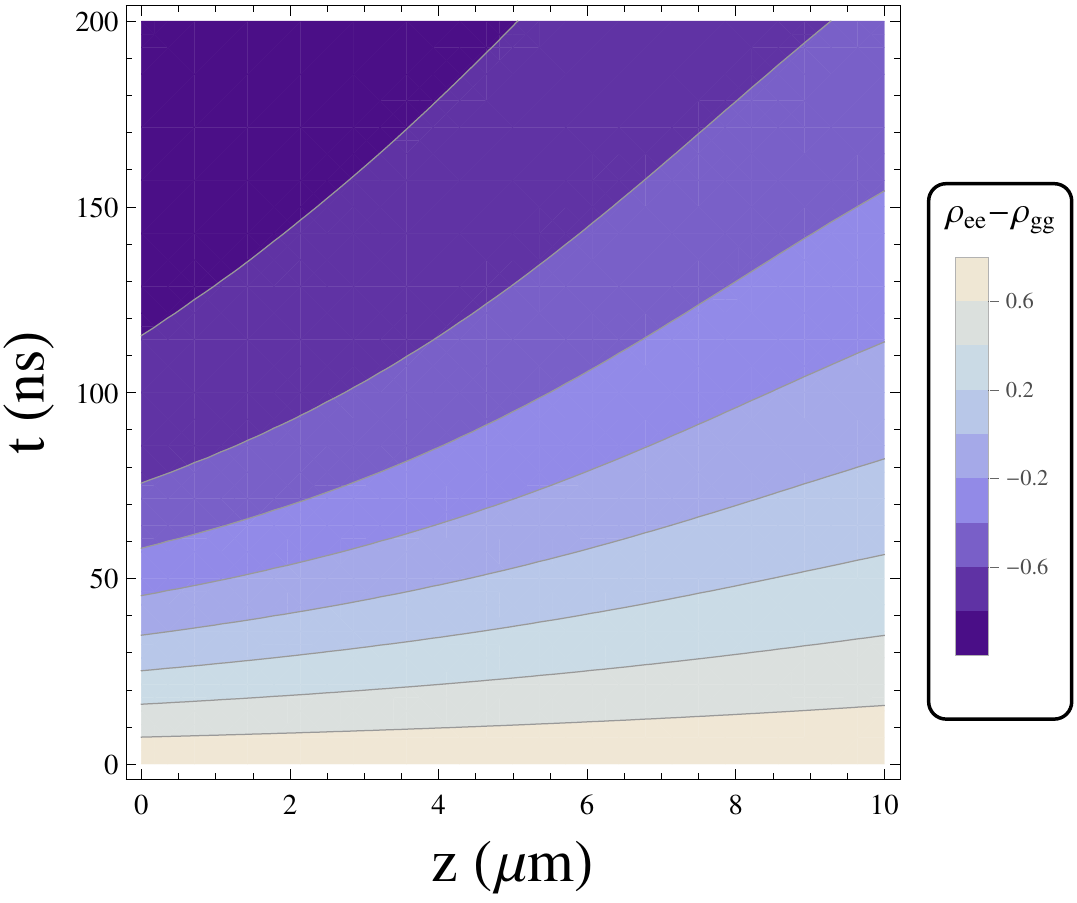}
\caption{\label{fig5} Population inversion $(\rho_{ee}-\rho_{gg})$ produced solely by the strong pulse as a function of time (here $t=0$ denotes the center of the strong pulse entering the sample) and position $z$ in the sample. }
\end{figure}

The strong pulse passes the nuclear sample prior to the weak pulse. This situation is depicted in Fig. \ref{fig2}. We see that in this case the   weak pulse signal can be suppressed by several orders of magnitude depending on the delay time $\Delta\tau$. The underlying mechanism for this suppression relies on two aspects: (i) the diminished nuclear ground state population left available for the later arriving weaker pulse and (ii) the building up of the weak pulse coherence term $\rho_{egw}$. 
The strong pulse produces a significant population of the excited states at $t=0$ and the population dynamics is still ongoing by the time the weaker pulse reaches the sample. This is illustrated  in Fig. \ref{fig4} where the contour plot of the time-dependent excited state population produced by the strong pulse as a function of position in the sample $z$ is presented. We see that at $t=10$ ns and $t=50$ ns after the passing of the strong pulse, a  still large amount of excitation is present in the sample and correspondingly fewer ground states are available for excitation by the weak pulse. However, this does not direcly explain why the arrival of the weak pulse with 50 ns delay time leads to a  more suppressed signal in Fig. \ref{fig2}(b) than the case of 10 ns delay, since the excited state population is  higher in the latter case. A study of the MBE for the two counter-propagating pulses (\ref{mbcounter}) reveals in the equation for the coherence $\rho_{egw}$ that it is the population inversion $(\rho_{ee}-\rho_{gg}
)$ which is decisive for the intensity of the scattered signal. Indeed, the weak pulse itself can produce only a weak excitation such that $\rho_{ee}-\rho_{gg}\simeq -1$. The  imaginary part of the coherence at $t=0$ is then given by the product between the incident (here Gaussian) pulse and the difference $(\rho_{ee}-\rho_{gg})$. However, with the strong pulse arriving prior to the weak pulse, the nuclear  population is first pumped in the excited state and $(\rho_{ee}-\rho_{gg})$ changes sign. A contour plot of the population inversion produced by the strong pulse is presented in Fig. \ref{fig5}. At $t=10$ ns when the weak pulse reaches the sample, the population inversion is approx. 0.8, leading to a smaller absolute value of the imaginary part of the initial coherence $\rho_{egw}$ for the weak pulse and a suppressed signal. If the weak pulse arrival is delayed up to 50 ns, the population inversion cancels with $\rho_{ee}-\rho_{gg}\simeq 0$ over most of the sample. The coherence $\mathrm{Im}[\rho_{egw}]$ and 
consequently the weak pulse signal is even more strongly suppressed. We note that the change of sign for the coherence term at $t=0$ does not play a role here since it only affects the initial phase of the scattered electric field and not  its intensity.

%

\subsection{$\Delta \tau>0$}

\begin{figure}[htbp]
\centering
\includegraphics[width=0.4\textwidth]{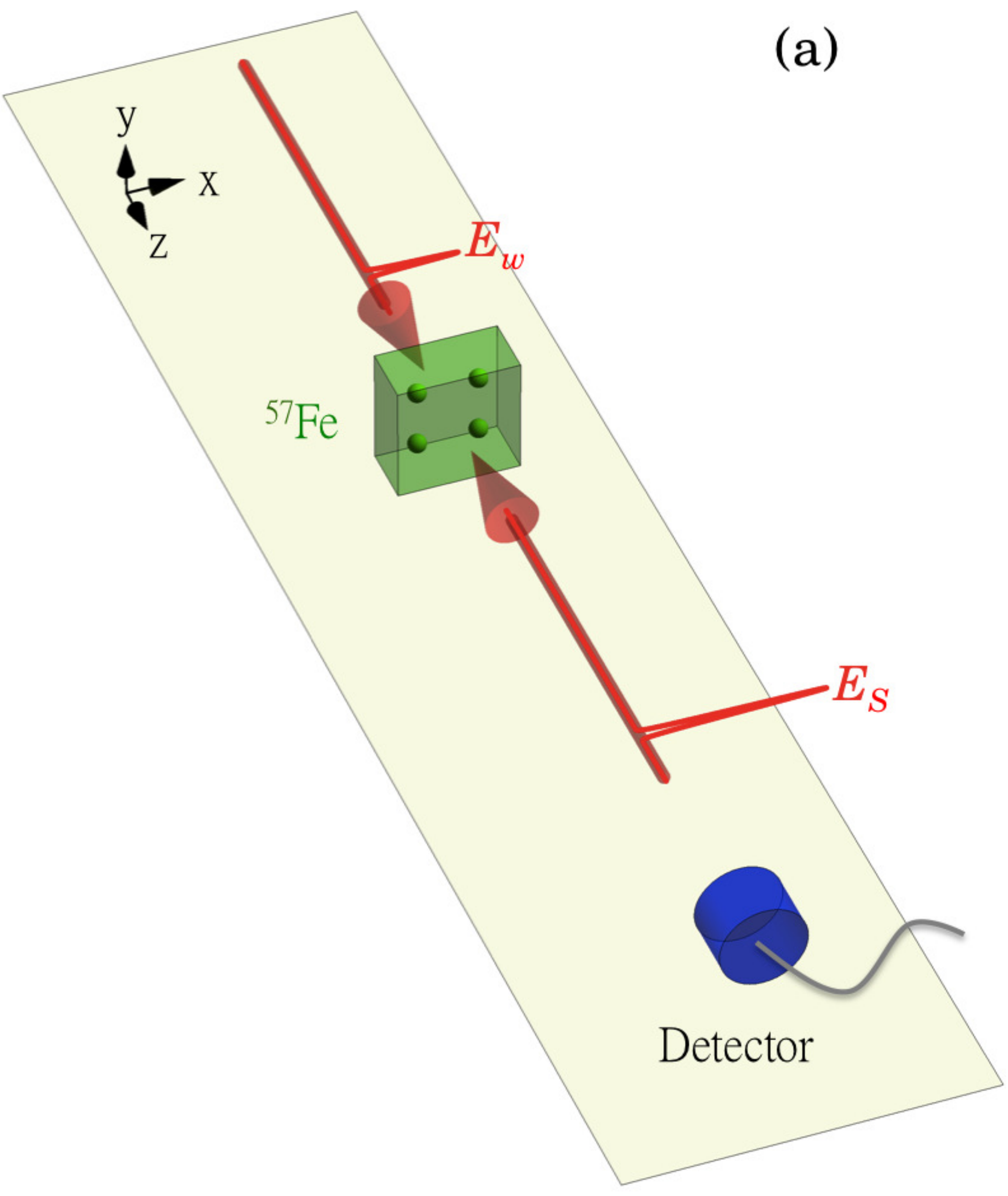}
\includegraphics[width=0.55\textwidth]{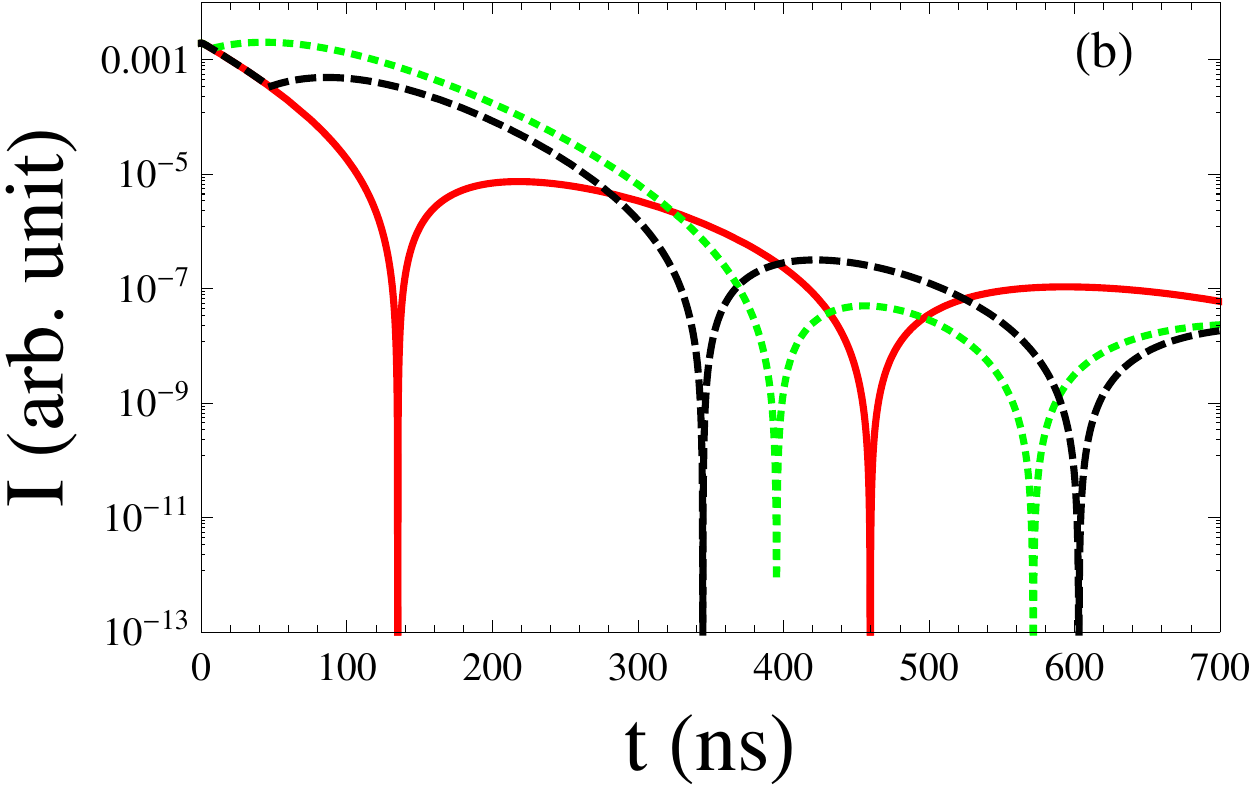}
\caption{\label{fig3} (a) Counter-propagating pulses setup with the weak pulse reaching the sample prior to the strong pulse. 
 (b)  NFS time spectra $I\propto|\Omega _{w}|^2$ for the weak pulse in the absence (red solid line) or presence of a stronger counter-propagating pulse $\Omega _{s}$. The weak pulse reaches the sample first and $\Delta\tau=10$ ns (green dotted line) and 50 ns (black dashed line). The time origin is again set by the center of the incident weak pulse reaching the sample at $z=0$. }
\end{figure}

The strong pulse arrives during the weak pulse propagation through the sample as shown in Fig. \ref{fig3}(a). Our results for this situation are depicted in Fig. \ref{fig3}(b). In this case the effect of the strong pulse arriving with a delay after the weak pulse is a substantial increase of the response of the latter. Similar arguments related to the strong-pulse-induced population inversion and coherence hold also in this case. However, the main difference now is that the weak pulse evolves first unperturbed and the coherence term
$\rho_{egw}$ is non-zero and decreasing when the strong pulse arrives. Thus, unlike in the previous situation discussed above, a sudden change in the sign of the population inversion will produce now an increase of $\rho_{egw}$ and consequently also an increase of the weaker pulse signal $|\Omega_{w}|^2$. The population inversion for both  $\Delta \tau=10$ ns and $\Delta \tau =50$ ns has similar values  leading to a comparable enhancements of the weak pulse signal for the green and the black curves in  \ref{fig3}(b). 
%


In order to further test our understanding of the two-pulse propagation dynamics in the nuclear sample, we have also considered a hypothetical  modified setup where the effect of the strong pulse on the excited state population for the weak pulse vanishes. The concrete example is a three-level $V$-type system where the two pulses each couple only to one of the two transitions, leading to the population of two different excited states.  The population inversion relevant for the weak pulse 
is therefore never changing sign, since  $\rho_{eew}\ll\rho_{gg}$ at all times. As expected, we observe the suppression of the weak pulse signal for all (positive and negative) delay times, with no enhancement observed.

\begin{figure}[htbp]
\centering
\includegraphics[width=0.6\textwidth]{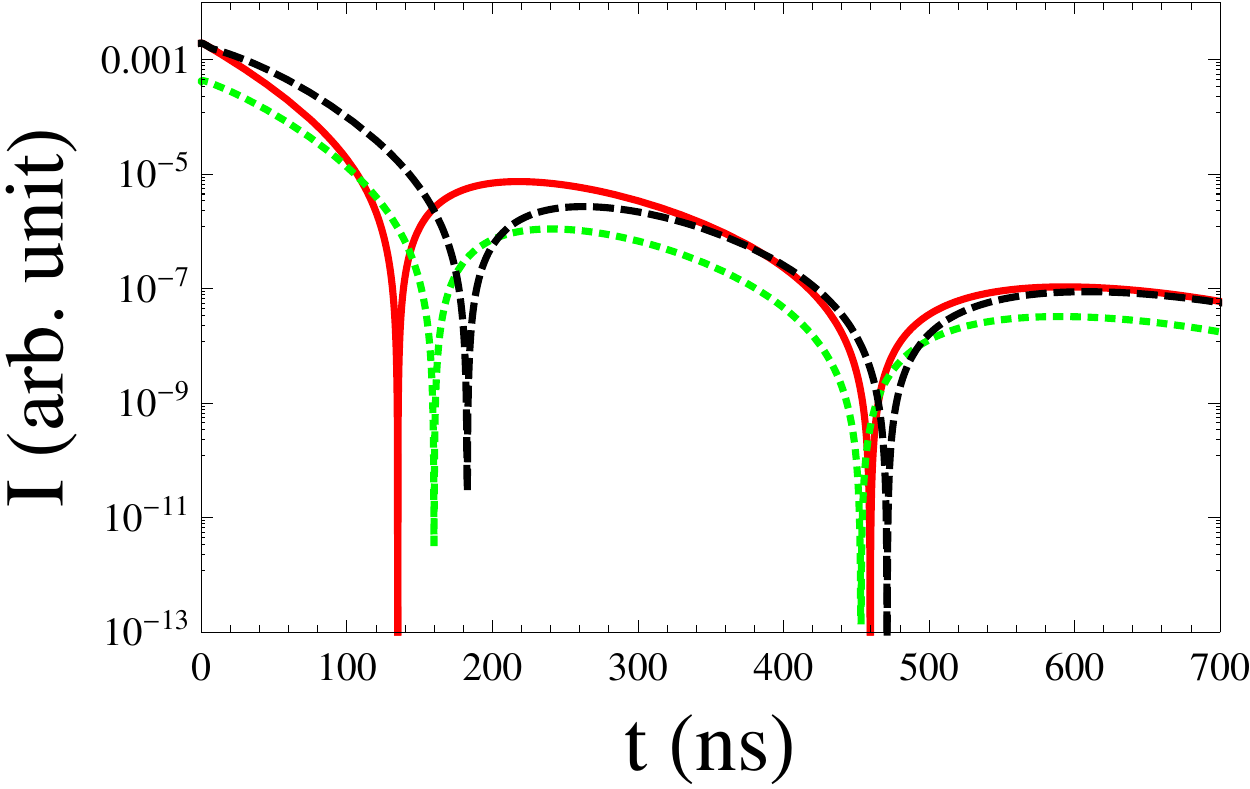}
\caption{\label{fig6} NFS spectra $|\Omega _{w}(L,t)|^2$ unperturbed (solid red line) and in the presence of a counter-propagating strong pulse with $\Omega _{s0}=100\Gamma$ reaching the sample with the pulse delay  $\Delta\tau=10$ ns (green dotted line) and $\Delta\tau=-10$ ns (black dashed line).}
\end{figure}

To summarize, prior arrival of a strong pulse can suppress  while a later arrival can enhance significantly the NFS signal of a weak pulse. This can have exciting applications in the framework of single-photon signal processing, for instance to enhance detection of single-photon wave packets. The key phenomenon here is the significant modification of the population inversion in the sample by the strong pulse. Obviously, in order to achieve the effects under investigation here, a certain intensity is required for  the strong pulse. The value assummed here of $\Omega _{s0}=200\Gamma$ corresponds to a peak intensity of $1.8\times 10^{22}$ W/cm$^2$, which is not far from present XFEL intensity values considering excellent focus \cite{mimura2009}. However, a  narrower bandwidth would be required which may be available only at future seeded XFEL facilities. For comparison, we present here our results also for a $\Omega _{s0}=100\Gamma$ for $\Delta \tau =\pm 10$ ns in Fig. \ref{fig6}. In this case, the suppression 
and enhancement effects are visible but already less spectacular. 

\section{Forwarding the nuclear response in time \label{magncontrol}}
\begin{figure}[htbp]
\centering
\includegraphics[width=0.6\textwidth]{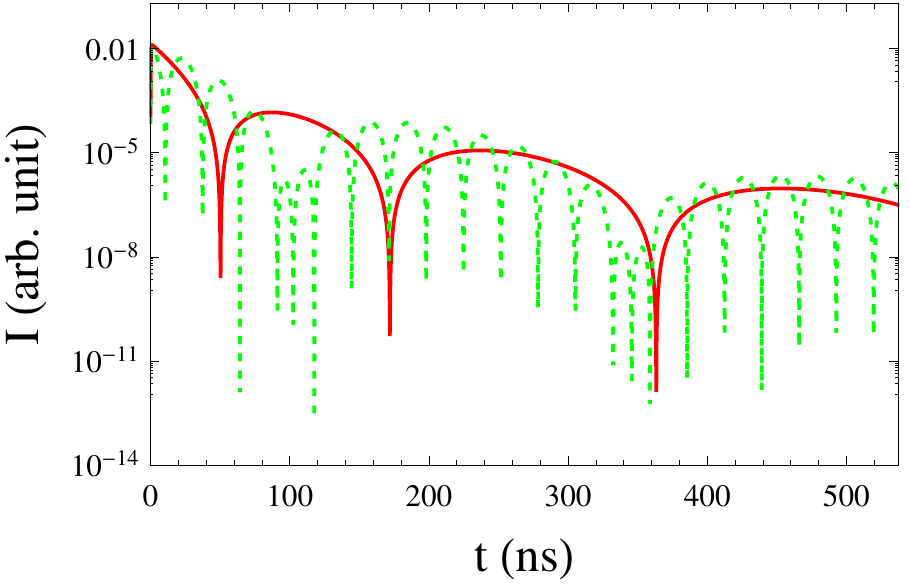}
\caption{\label{fig7}  Intensity of the coherent scattered light for a single nuclear transition in the absence of hyperfine splitting and for an effective thickness of  $\xi=40$ (solid red line). In the presence of the magnetic field, the two $\Delta m=0$ transitions interfere and introduce a quantum beat in the spectrum (green dotted line).}
\end{figure}

We now investigate the case when only one pulse propagates  resonantly through the sample, however under the action of a time-dependent magnetic field. In the absence of the magnetic field, the ${^{57}}\mathrm{Fe}$ nuclei behave as two-level systems. If the magnetic field is switched on, the introduced hyperfine splitting renders six transitions possible. We consider in the following a setup for which the incident pulse polarization and the geometry of the magnetic field, when present, allow only for the driving of the two $\Delta m=0$ transitions. As further parameters, the magnetic field intensity of $B$=17.2 T and  an effective thickness for the two-level nuclear system of $\xi=40$ are envisaged. The hyperfine splitting effectively produces in this case  a shift to a smaller value of $\xi$ since the ground state population distributes half-half over the two hyperfine-split ground states with $m_g=-1/2$ and $m_g=1/2$. This is  illustrated by the shapes of the dynamical beat in the NFS time spectra for the 
two cases in the presence and absence of the magnetic field presented in Fig. \ref{fig7}. The envelope of the quantum beat follows here the dynamical beat corresponding to $\xi=20$.

\begin{figure}[htbp]
\begin{center}
\includegraphics[width=\textwidth]{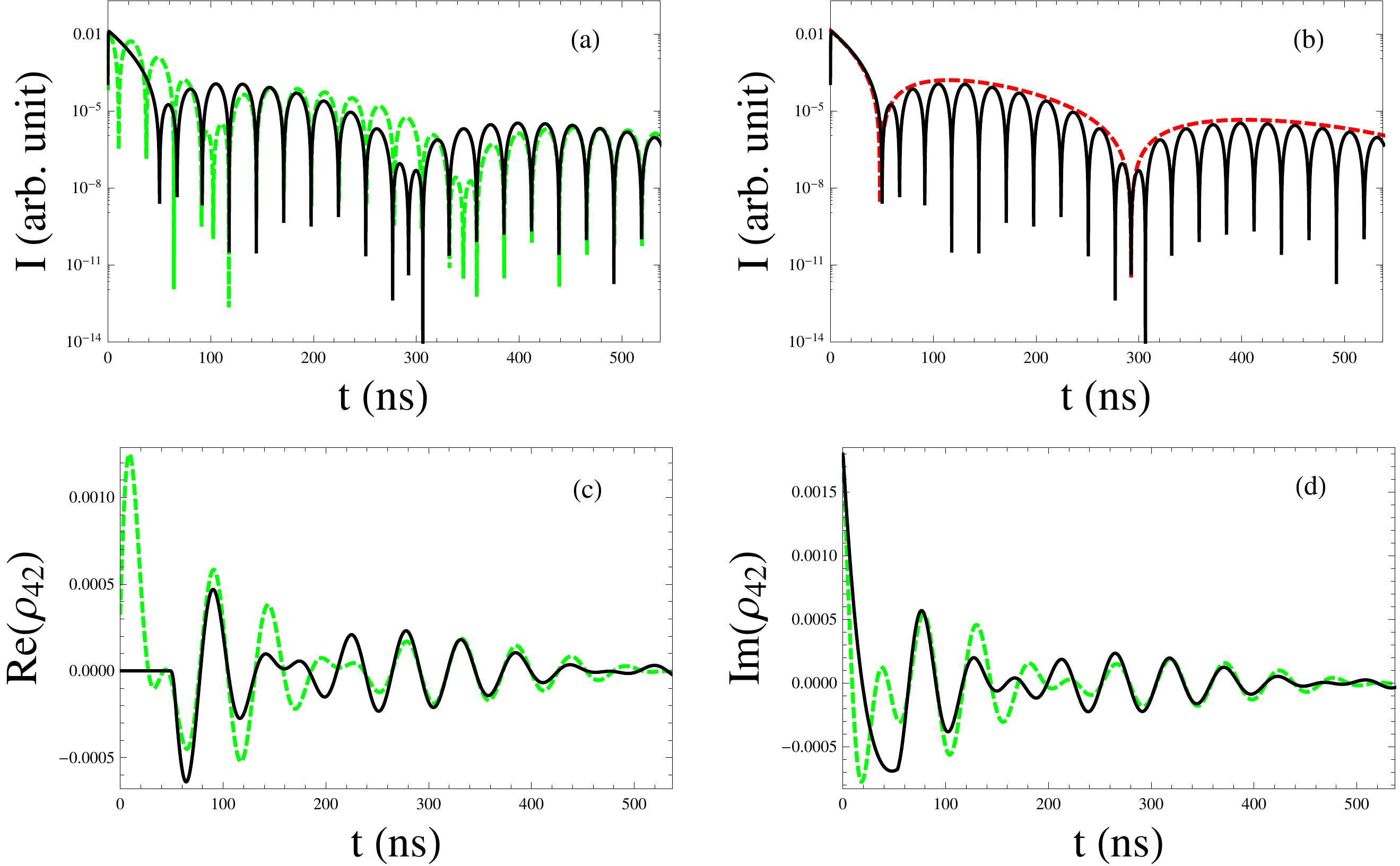}
\caption{\label{fig8}  NFS time spectra (a,b) and the real (c) and imaginary (d) parts of the coherence term $\rho_{42}$. The dashed green line depicts the case of scattering in the presence of a magnetic field at all times, while the black  line presents the case of the magnetic field being switched on rapidly at $t=50$ ns.  Correspondingly a 50 ns shift of the signal can be observed. The red solid line in (b) illustrates for comparison the dynamical beat envelope for $\xi=20$ as discussed in the text. } 
\end{center}
\end{figure}

We now attempt to switch between the degenerate and non-degenerate nuclear level systems by turning the magnetic field on or off. Coherent storage of nuclear excitation has been theoretically shown  to be possible when the magnetic field present at $t=0$ when the incident SR or XFEL pulse arrived is switched off at certain times. A by-product of the coherent storage is that the NFS signal appears to be shifted backwards in time. Here, we investigate the opposite situation. Initially, the incident pulse hits the ${^{57}}\mathrm{Fe}$ sample in the absence of any hyperfine magnetic field. The magnetic field is switched on later, in our first example at $t_0=50$ ns, when the minimum of the dynamical beat is reached. Quantum beats then appear in the NFS spectrum as a result of the two hyperfine transitions that can constructively or destructively interfere. This situation is illustrated in Fig. \ref{fig8}(a) by the black  line. The signal for $t<50$ ns can be described by  $\xi [J_1(2\sqrt{\xi\Gamma t})]^2 e^{-\Gamma t}/(\Gamma t)$ where $\xi=40$. Later on, after the hyperfine magnetic field has been switched on, the  envelope  illustrated in  \ref{fig8}(b) by the red curve can be described as $\xi'[J_1(2\sqrt{\xi'\Gamma (t+t_0)})]^2 e^{-\Gamma t}/(\Gamma (t+t_0))$ where $\xi'=\xi/2$. The comparison between the case with magnetic field at all times and magnetic field only after $t=50$ ns is presented for the NFS spectra and the real and imaginary parts of the coherence term $\rho_{42}$ in Figs. \ref{fig8}(a), (c) and (d). 

\begin{figure}
\begin{center}
\includegraphics[width=0.6\textwidth]{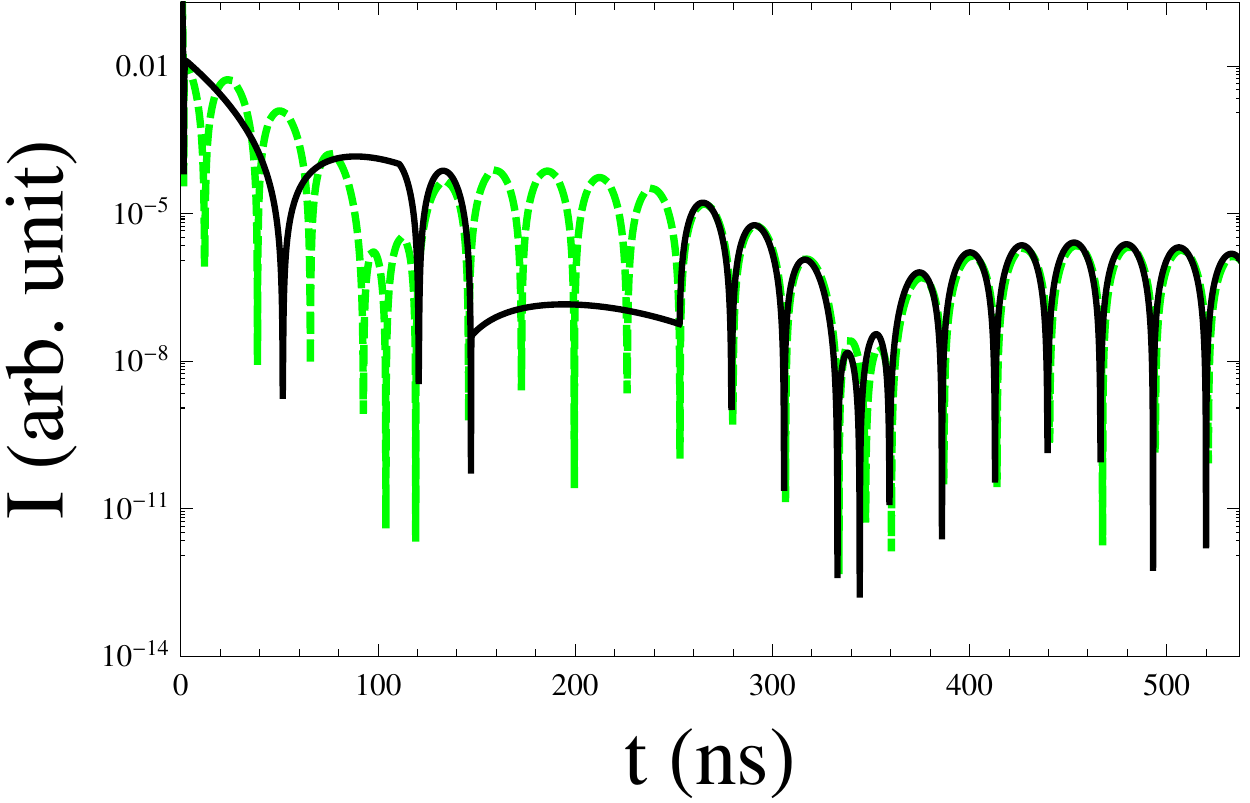}
\caption{\label{fig9} NFS time spectra for a hyperfine splitting constant in time (green dashed line) compared to the switching scheme turning the magnetic field on ($t_1=105$ ns), off ($t_2=145$ ns), and on again ($t_3=251$ ns),  described in the text (black  line). }
\end{center}
\end{figure}

The surprising feature of the two NFS spectra in the presence of magnetic field in Figs. \ref{fig8} is that the system dynamics, including both the scattered  signal and the coherence terms, is identical and just {\it shifted in time} up to the effect of the  exponential spontaneous decay. Indeed, a numerical comparison of the unperturbed and shifted spectra shows that they coincide when considering the 50 ns time shift and accounting for the corresponding spontaneous decay.
 The turning on of the magnetic field after the incident radiation pulse arrived thus displaces the signal forward by the same time interval $\Delta t$=50 ns compared to the spectrum with constant hyperfine splitting.  This is the opposite effect compared to the coherent photon storage presented in Ref. \cite{Liao2012a}. In order to demonstrate this, we design a succession of four manipulations on the magnetic field in order to produce the forward shift of the signal and the coherent storage. The results are illustrated in Fig. \ref{fig9}.  The incident pulse reaches the nuclear sample at $t_0=0$ when there is no magnetic field present and no hyperfine splitting in the sample. Later on, at $t_1=105$ ns, the magnetic field is switched on rapidly and the quantum beats occur. At a later time, when a quantum beat minimum is reached ($t_2=145$ ns), the magnetic field is switched off again and coherent storage \cite{
Liao2012a} is achieved. The effect of the  coherent storage is to shift now the signal backwards, i.e., towards longer scattering times, thus canceling the effect of the first signal shift forward in time. Finally, at $t_3=251$ ns the magnetic field is switched on and we retrieve the NFS signal which matches exactly the situation when the magnetic field was on during the whole scattering period, as shown in Fig. \ref{fig9}. The shifts forwards and backwards in time cancel each other since $t_0-t_1\simeq t_3-t_2$. We would like to emphasize here that, just as in the case of coherent photon storage \cite{Liao2012a}, shifting the signal forwards in time occurs preserving the phase information, i.e., we witness the phase-sensitive shift of the signal in time. 

\begin{figure}
\begin{center}
\includegraphics[width=0.8\textwidth]{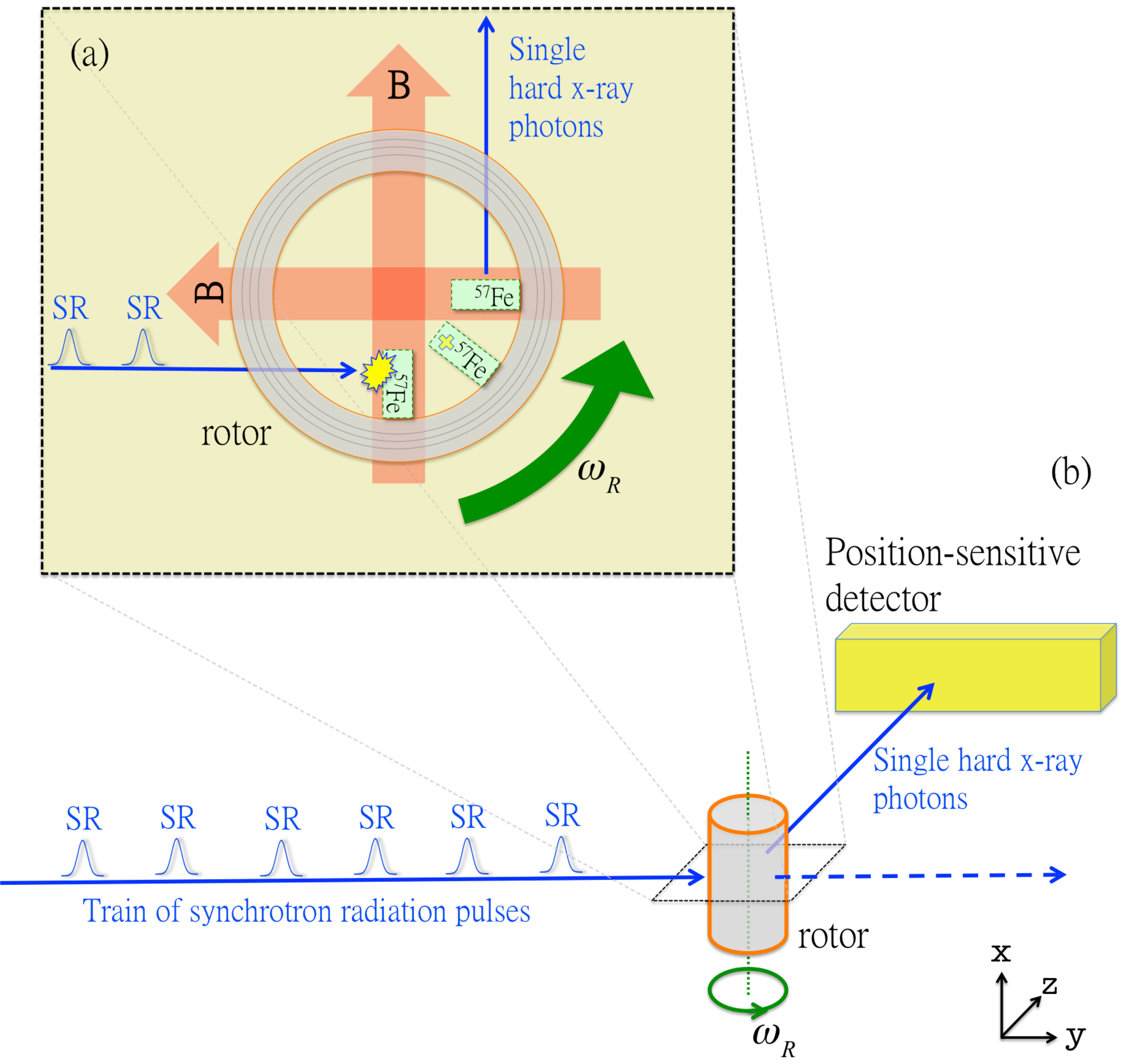}
\caption{\label{fig10} Sketch of the lighthouse setup for the coherent storage of hard x-ray single photons. (a)
Bird view of the lighthouse setup. Gray area depicts the side view of the rotor rotating with angular
frequency $R$, the two red wide arrows illustrate the regions with confined static magnetic field {\bf B}
and the blue arrows the trajectories of SR and emitted single hard x-ray photons. The light green
rectangles depict snapshots of the rotating $^{57}$Fe target attached on the inner surface of the rotor.
(b) The geometric arrangement of the lighthouse scheme.}
\end{center}
\end{figure}

Thus, temporal signal control can be achieved via fast switching on and off of the magnetic field. The experimental challenges for the control on ns time scale of strong magnetic fields have been first addressed in Ref. \cite{Liao2012a}. 
The most promising solution involves a material  with no intrinsic nuclear Zeeman splitting like stainless steel Fe$_{55}$Cr$_{25}$Ni$_{20}$ \cite{Smirnov1996NE, jex1997}. The  challenge is to turn off and on the external magnetic fields of few Tesla on the ns time scale. According to the calculations presented in Ref.~\cite{Liao2012a}, the raising time of the {\bf B} field should be shorter than 50 ns (the raising time was considered 4 ns for all presented cases). This could be achieved by using small single- or few-turn coils and a moderate pulse current of approx. 15 kA from low-inductive high-voltage ``snapper'' capacitors \cite{Miura2003}. Another mechanical solution, e.g., the lighthouse setup \cite{Roehlsberger2000} could be used to move the excited target out of and into a region with confined static {\bf B} field. The nuclear lighthouse setup  is based on a rotating sample.  This changes the direction of the coherently emitted photon which is always in the forward direction with respect to the 
sample, thus explaning the name ``lighthouse effect''. The rotation can be used to bring the sample in and outside a region with confined static magnetic field. The switching time is then given by the time needed for the rotation of the sample from the edge of the confined magnetic field region to the outside, magnetic-field free region. 
With the setup illustrated in Fig. \ref{fig10}, we estimate that a rotor with rotational frequencies $R$ of up to 70 kHz and a diameter of few mm \cite{Roehlsberger2000} is fast enough to rotate the sample 
out a depth of few $\mu$m in a few tens of ns. If mastered, this fast magnetic-field switching  would allow elaborated coherent control over the nuclear excitation in NFS and accordingly over the dynamics of single x-ray photon wave packets. 

\section{Conclusions \label{Conclusions}}
Nuclei, although typically difficult to drive with electromagnetic fields, may be the key to coherently control single x-ray photons in a NFS setup. Accordingly, means to coherently control the x-ray quanta may have a great potential  for exciting quantum applications. Here, we have investigated theoretically two advanced field-control schemes to enhance, suppress or shift in time the single x-ray photon signal. Our theoretical approach relies on the MBE, which are computationally advantageous and allow the straightforward treatment of time-dependent perturbations in the resonant propagation of light through the nuclear medium. We have shown that the simultaneous propagation of two pulses through the same nuclear sample can lead to the transfer of signal intensity between the two, depending on the corresponding intensities and time delay between the pulses. Thus, the presence of a strong pulse, for instance produced by the XFEL, can lead to the enhancement or suppression of the signal of a weaker excitation, potentially comprising a single resonant x-ray photon. Furthermore, the signal of such a weak excitation can be shifted forward in time by the alternation between scattering intervals  in the presence and absence of a hyperfine  magnetic field. This is the inverse effect of coherent photon storage and may become a valuable technique if single x-ray photons are to become the information carriers in future photonic devices. 

\ack
AP gratefully acknowledges the hospitality of the University of Strathclyde in the framework of the Scottish Universities Physics Alliance visitor program.

\appendix
\section{The MBE for circular polarization}
A circularly polarized incident pulse will drive either the   $\Delta m=m_e-m_g=1$ (this field is denoted below by $\Omega^+$) or the  $\Delta m=m_e-m_g=-1$ (respectively $\Omega^-$) transitions between the two ground state and four excited state hyperfine levels. Using the  level notations defined in the text in Sec. \ref{mbe}, we obtain
the Bloch equations
\begin{eqnarray}
\partial _t\rho _{11}&=&\Gamma(C_{14}^2\rho _{44}+C_{15}^2\rho _{55}+C_{16}^2\rho _{66})
\nonumber \\
&-&\frac{i}{2}[C_{14}(\Omega_{p}^-\rho _{14}-\mathrm{c.c.})+C_{16}(\Omega_{p}^+\rho _{16}-\mathrm{c.c.})]\,  , \nonumber \\
\partial _t\rho _{22}&=&\Gamma(C_{23}^2\rho _{33}+C_{24}^2\rho _{44}+C_{25}^2\rho _{55})
\nonumber \\
&-&\frac{i}{2}[C_{23}(\Omega_{p}^-\rho _{23}-\mathrm{c.c.})+C_{25}(\Omega_{p}^+\rho _{25}-\mathrm{c.c.})]\,  , \nonumber \\
\partial _t\rho _{32}&=&-\frac{1}{2}(2i\Delta_{p,3\rightarrow2}+C_{23}^2\Gamma)\rho _{32}
-\frac{i}{2}C_{23}\Omega_{p}^-(\rho_{33}-\rho _{22})-\frac{i}{2}C_{25}\Omega_{p}^+\rho _{35}\,  , \nonumber \\
\partial _t\rho _{33}&=&-C_{23}^2\Gamma\rho _{33}
+\frac{i}{2}C_{23}(\Omega_{p}^-\rho _{23}-\mathrm{c.c.})\,  , \nonumber \\
\partial _t\rho _{41}&=&-\frac{1}{2}(2i\Delta_{p,4\rightarrow1}+C_{14}^2\Gamma+C_{24}^2\Gamma)\rho _{41}
-\frac{i}{2}C_{14}\Omega_{p}^-(\rho_{44}-\rho _{11})-\frac{i}{2}C_{16}\Omega_{p}^+\rho _{46}\,  , \nonumber \\
\partial _t\rho _{44}&=&-(C_{14}^2+C_{24}^2)\Gamma\rho _{44}
+\frac{i}{2}C_{14}(\Omega_{p}^-\rho _{14}-\mathrm{c.c.})\,  , \nonumber \\
\partial _t\rho _{52}&=&-\frac{1}{2}(2i\Delta_{p,5\rightarrow2}+C_{15}^2\Gamma+C_{25}^2\Gamma)\rho _{52}
-\frac{i}{2}C_{25}\Omega_{p}^+(\rho_{55}-\rho _{22})-\frac{i}{2}C_{23}\Omega_{p}^-\rho _{53}\,  , \nonumber \\
\partial _t\rho _{53}&=&-\frac{1}{2}(2i\Delta_{p,5\rightarrow2}-2i\Delta_{p,3\rightarrow2}+C_{15}^2\Gamma+C_{25}^2\Gamma+C_{23}^2\Gamma)\rho _{52}
\nonumber \\
&-&\frac{i}{2}C_{23}\Omega_{p}^-\rho_{52}+\frac{i}{2}C_{25}\Omega_{p}^+\rho _{23}\,  , \nonumber \\
\partial _t\rho _{55}&=&-(C_{15}^2+C_{25}^2)\Gamma\rho _{55}
+\frac{i}{2}C_{25}(\Omega_{p}^+\rho _{25}-\mathrm{c.c.})\,  , \nonumber \\
\partial _t\rho _{61}&=&-\frac{1}{2}(2i\Delta_{p,6\rightarrow1}+C_{16}^2\Gamma)\rho _{32}
-\frac{i}{2}C_{16}\Omega_{p}^+(\rho_{66}-\rho _{11})-\frac{i}{2}C_{14}\Omega_{p}^-\rho _{64}\,  , \nonumber \\
\partial _t\rho _{64}&=&-\frac{1}{2}(2i\Delta_{p,6\rightarrow1}-2i\Delta_{p,4\rightarrow1}+C_{14}^2\Gamma+C_{16}^2\Gamma+C_{24}^2\Gamma)\rho _{64}
\nonumber \\
&-&\frac{i}{2}C_{14}\Omega_{p}^-\rho_{61}+\frac{i}{2}C_{16}\Omega_{p}^+\rho _{14}\,  , \nonumber \\
\partial _t\rho _{66}&=&-C_{16}^2\Gamma\rho _{66}
+\frac{i}{2}C_{16}(\Omega_{p}^+\rho _{16}-\mathrm{c.c.})\,  , \nonumber \\
\end{eqnarray}
with the Maxwell equations for the Rabi frequencies $\Omega_p^+$ and $\Omega_p^-$ of the two components given by
\begin{eqnarray}
\frac{1}{c}\partial _t\Omega_{p}^++\partial _z\Omega_p^+&=&i\frac{\eta}{2}\left(\frac{\rho _{61}}{C_{16}}+\frac{\rho _{52}}{C_{25}}\right)\, , \nonumber \\
\frac{1}{c}\partial _t\Omega_{p}^-+\partial _z\Omega_p^{-}&=&i\frac{\eta}{2}\left(\frac{\rho _{41}}{C_{41}}+\frac{\rho _{32}}{C_{23}}\right)\, .
\end{eqnarray}
%

\section*{References}

\bibliographystyle{iopart-num}
\bibliography{nfs}
\end{document}